\documentclass[conference]{IEEEtran}
\usepackage[utf8]{inputenc}
\usepackage{booktabs}
\usepackage{subcaption}
\usepackage{algorithm}
\usepackage{algpseudocode}
\usepackage{tikz}
\usetikzlibrary{arrows.meta,positioning,shapes,calc}
\usepackage{pgfplots}
\pgfplotsset{compat=1.18}
\usepackage{cite}
\usepackage{amsmath,amssymb,amsfonts}
\usepackage{graphicx}
\usepackage[inkscapelatex=false]{svg}
\usepackage{tabularx}
\usepackage{textcomp}
\usepackage{xcolor}
\usepackage{multirow}
\usepackage{array}  
\usepackage{float}
\usepackage{hyperref}
\usepackage{xcolor}
\usepackage{orcidlink}
\hypersetup{
    colorlinks=true,
    linkcolor=blue,
    citecolor=blue,
    filecolor=magenta,      
    urlcolor=cyan,
    pdftitle={},
    pdfpagemode=FullScreen,
}

\begin{document}

\title{Edge-Based QoS-Aware Adaptive Task Placement: A Closed-Loop Control in Multi-Robot Systems}

\author{\IEEEauthorblockN{
    Thien Tran\IEEEauthorrefmark{1},
    Jonathan Kua\IEEEauthorrefmark{1},
    Thuong Hoang\IEEEauthorrefmark{1},
    Minh Tran\IEEEauthorrefmark{2},
    Honghao Lyu\IEEEauthorrefmark{3} and
    Jiong Jin\IEEEauthorrefmark{4}
    } \\
    \IEEEauthorrefmark{1}Deakin University, Australia;
    \IEEEauthorrefmark{2}RMIT University, Vietnam;
    \IEEEauthorrefmark{3}Zhejiang University, China; \\
    \IEEEauthorrefmark{4}Swinburne University of Technology, Australia; \\
    {\{peter.tran, jonathan.kua, thuong.hoang\}@deakin.edu.au}; {minh.tranquang@rmit.edu.vn}; \\
    {lvhonghao@zju.edu.cn};
    {jiongjin@swin.edu.au} \\
}

\maketitle
\begin{abstract}
Multi-robot systems (MRS) increasingly offload compute-intensive perception tasks to edge nodes to meet strict time-sensitive Quality-of-Service (QoS) constraints. However, static task orchestration on a shared edge node can severely degrade QoS due to network latency, jitter, and edge-resource contention. We present a pilot edge-centric MRS testbed using Raspberry Pi nodes to evaluate a camera-to-manipulator pipeline under three modes: local execution, static offloading, and a QoS-aware Adaptive Task Placement (ATP) controller. ATP scores candidate placements using a multi-metric cost (normalized latency, CPU utilization, and switching overhead) over two-second control windows. The closed-loop visual servoing testbed is instrumented with sub-millisecond clock synchronization, network emulation, and detailed monitoring of multiple metrics across nodes to capture realistic jitter. Experimental results under compute-stress and network-fault scenarios show that static edge offloading reduces on-board CPU load but amplifies tail latency and deadline misses. In contrast, the QoS-aware ATP controller, by switching task placement based on measured latency and utilization thresholds, consistently lowers deadline violations and tail latency. Overall, the results position ATP as a practical edge-side control primitive for MRS and concrete design guidelines for Cloud–Edge Robotics deployments within the broader cloud-fog automation, while motivating QoS-aware multi-objective workload orchestration for industrial~cyber–physical~systems.
\end{abstract}


%=====================
\section{Introduction}
%=====================
% Para #1
Multi-robot systems (MRS) are increasingly deployed in complex industrial environments, where fleets of robots share perception, manipulation, inspection, and navigation tasks as part of larger industrial cyber--physical systems (ICPS)~\cite{jin_cfa_jsac25, jin_cfa_tii24}. In these settings, robots are expected to run compute-intensive workloads while operating under tight timing, safety, and quality-of-service (QoS) constraints~\cite{r_andreoli_multi-domain_2025}. Cloud and edge robotics have emerged as complementary approaches to push heavy computation off-board, allowing robots to tap into more powerful compute and shared models while remaining relatively low-cost and resource-constrained on-board~\cite{w_chen_review_2024}. Mobile edge computing (MEC) and edge-enabled robotics surveys~\cite{dong_task_2024, tahir_edge_2025} highlight latency reduction, bandwidth savings, and improved autonomy as key drivers for moving computation from distant cloud to nearby edge following Cloud-Edge-Robotics (CER)~\cite{tran_cftel_indin25, tran_usv_indin25} within the broader cloud-fog automation (CFA).

% Para #2
A large body of work has explored task offloading and resource management in MEC, and several cloud/edge robotics prototypes demonstrate remote execution of perception, mapping, and planning components ~\cite{y_mao_survey_2017, chinchali_network_2021}. In robotics, edge-accelerated visual odometry and multi-robot SLAM show that offloading perception and mapping to edge nodes can reduce latency and improve perception quality~\cite{l_morra_mixo_2023}, while edge-enabled digital twins for multi-robot collision avoidance and teleoperation demonstrate that edge computation enables low-latency feedback~\cite{mtowe_low-latency_2025, coltran_jii_25, iotvr_jamt_25}. However, these benefits are inherently fragile: congestion on shared links, variable delay and jitter, and contention on edge nodes can erode QoS, while many existing studies rely on simulation or single-shot tasks and report average latency, even though practical deployments must sustain closed-loop behavior under varying conditions.

% Para #3
Hence, there remains limited empirical evidence from MRS testbeds that simultaneously (i) instrument one-way control-loop latency, (ii) inject realistic and controllable network impairments, and (iii) evaluate a QoS-aware adaptive task placement (ATP) policy by utilizing the existing studies. This paper addresses this gap by constructing a compact edge-centric testbed hosting a camera-driven manipulator task representative of industrial pick-and-place or tracking scenarios. The testbed implements a sensing–perception–planning–control pipeline and supports three execution modes: local execution, static offloading, and a QoS-aware ATP controller. In summary, this study makes the following contributions in the field:
\begin{itemize}
    \item \textbf{Edge-centric MRS testbed:} Implements a reproducible multi-robot testbed with an edge node for closed-loop QoS evaluation under controllable network conditions.
    \item \textbf{QoS-aware ATP controller:} Designs a lightweight ATP controller that chooses between local and edge execution for perception and planning based on the cost metrics.
    \item \textbf{Empirical comparison study:} Experimentally compares fully local, static edge offloading, and ATP-driven execution under compute-stress and network-fault scenarios.
\end{itemize}

% Para #4
The remainder of the paper is as follows. Section~\ref{sec2} describes the system architecture. Section~\ref{sec3} defines the execution modes and the QoS model. Section~\ref{sec4} presents the experimental setup and evaluation. Section~\ref{sec5} discusses the results. Section~\ref{sec6} concludes the study and outlines~future~work.

%======================================
\newpage
\section{MRS Testbed and System Overview}\label{sec2}
%======================================
\subsection{Hardware and Network Architecture}
\begin{figure}[t]
    \centering
    \includegraphics[width=\linewidth]{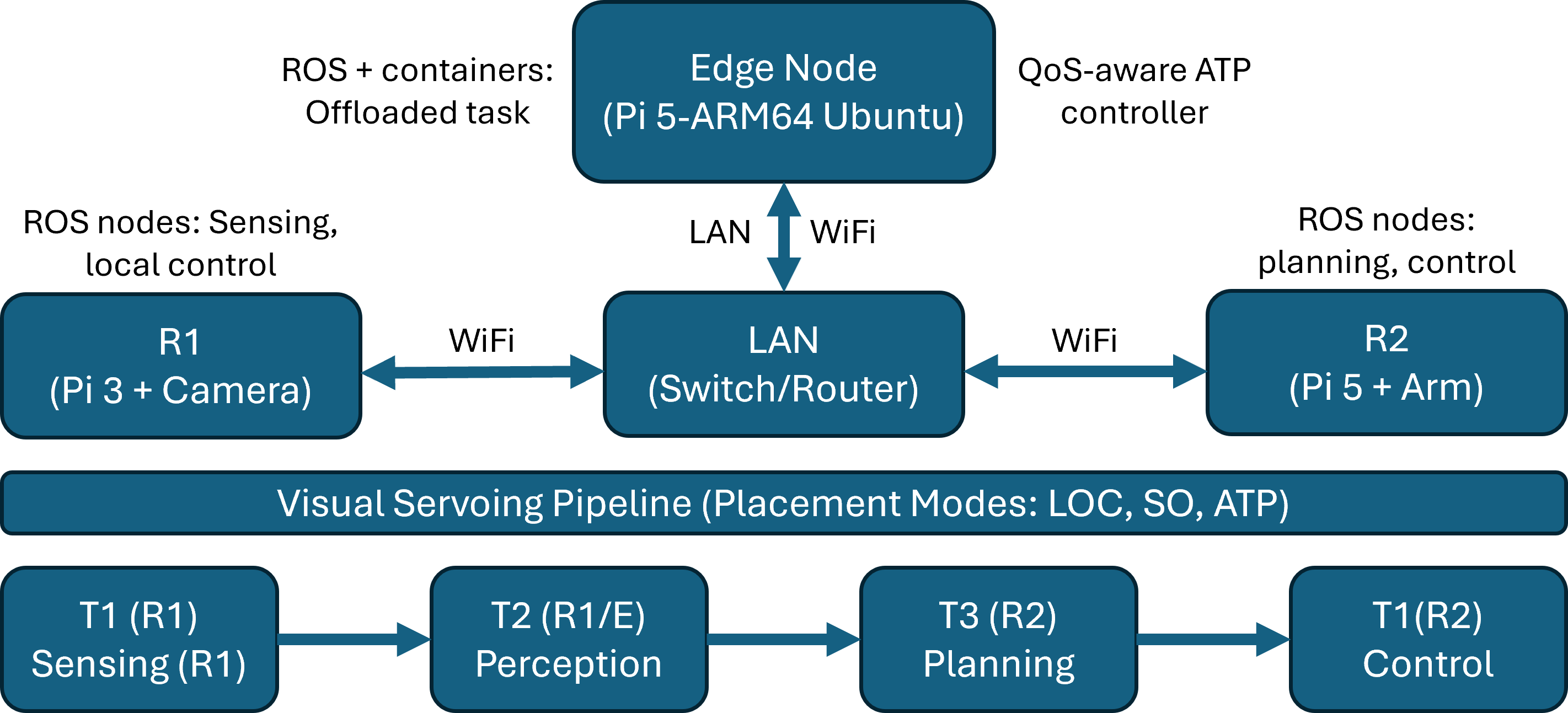}
    \caption{Edge-centric MRS testbed architecture and logical sensing--perception--planning--control pipeline.}
    \label{fig:arch}
\end{figure}

The testbed comprises two heterogeneous robot nodes and a single edge node interconnected through a local area network (LAN), with an additional wide-area (WAN) uplink from the edge to an external network for remote monitoring, as shown in Fig.~\ref{fig:arch}. All intra-testbed nodes are time-synchronized and primarily connected via WiFi through a single switch/router to minimize uncontrolled variability during experiments.

\begin{itemize}
    \item \textbf{Robot 1 (R1, ``light'' node):} Raspberry Pi~3B equipped with a monocular camera and a simple base, which is responsible for image acquisition and hosts lightweight local processing and control tasks.
    \item \textbf{Robot 2 (R2, ``heavy'' node):} Raspberry Pi~5 driving a small robotic arm, which executes motion control based on high-level pose and joint commands received from upstream perception and planning components.
    \item \textbf{Edge node (E):} An ARM64 Pi--based Ubuntu edge server connected to the same LAN, which runs offloaded perception tasks when the system operates in static edge offload or QoS-aware ATP modes.
    \item \textbf{Network:} R1, R2, and E are interconnected in the LAN switch/router via WiFi. Baseline round-trip time (RTT) and throughput are characterized before experimentation, and additional delay and jitter are injected using software network emulation on the edge-side interface to realize stressed network scenarios.
\end{itemize}

\begin{table}[b]
  \centering
  \caption{Testbed hardware and software configuration.}
  \label{tab:testbed}
  \begin{tabular}{@{}lll@{}}
    \toprule
    Node & Hardware & Software \\ \midrule
    R1 (light) & Raspberry Pi 3B, 1 GB RAM & Raspbian, ROS 2 \\
    R2 (heavy) & Raspberry Pi 5, 8 GB RAM & Raspbian, ROS 2 \\
    Edge (E) & Raspberry Pi 5, 16 GB RAM & Ubuntu, Docker, ROS 2 \\
    Network & WiFi LAN (802.11ac/ax) & Baseline RTT $\approx$ 1--2 ms \\
    Control loop & Period ($P$) 20--50 ms & Deadline $D \leq P$ \\ \bottomrule
  \end{tabular}
\end{table}

Fig.~\ref{fig:arch} highlights the edge-centric topology and the mapping between edge-robot nodes, the wireless LAN infrastructure, and the visual servoing pipeline across execution modes.

While this pilot testbed utilizes accessible Commercial Off-The-Shelf (COTS) single-board computers to ensure open-source reproducibility, the architecture directly maps to standard ICPS deployment models. Specifically, the resource-constrained R1 node emulates either a legacy industrial sensor gateway or a smart camera with limited local compute resources. The R2 node represents a modern ARM-based embedded Programmable Logic Controller (PLC) or a dedicated robot end-effector controller. Finally, the Ubuntu-based Edge Node serves as an on-premises Industrial PC (IPC) stationed on the factory floor, designed to offload compute-intensive machine vision workloads from the embedded PLCs.

% \begin{figure*}[t]

%======================================
\subsection{Workload and Task Pipeline}
%======================================
The demonstrator workload instantiates a camera-to-manipulator visual servoing loop representative of a simple industrial pick-and-place and target-tracking operation. Conceptually, a camera mounted on R1 observes the scene, while a small arm driven by R2 moves in response. The pipeline is decomposed into four main sequence tasks as follows:
\begin{itemize}
  \item \textbf{T1 -- Sensing:} Image frames are captured at a fixed rate.
  \item \textbf{T2 -- Perception:} A vision module processes each frame to extract target features, such as the target position.
  \item \textbf{T3 -- Planning/Decision:} A controller computes desired end-effector poses and manipulator-joint references.
  \item \textbf{T4 -- Control/Actuation:} The end-robot translates the references into joint commands and applies them to the manipulator.
\end{itemize}

The four tasks form a periodic closed loop, with one control cycle spanning from image acquisition in T1 to command execution in T4. T1, T3, and T4 are naturally anchored to R1 and R2, respectively, while T2 is experimentally relocatable: it can execute locally on the robots or be offloaded to the edge node, depending on the execution mode (local, static offload, or ATP). The resulting visual servoing pipeline and its mapping onto R1, E, and R2 are summarized in Fig.~\ref{fig:arch}.

%======================================
\subsection{Software Stack and Monitoring}
%======================================
All nodes run a Linux-based OS with ROS as the middleware for distributed communication. Tasks T1–T4 are implemented as ROS nodes that exchange messages via topics and services, while the edge uses Docker containerization to encapsulate offloaded components and streamline deployment and experimentation. A lightweight monitoring layer records (i) timestamps at key interfaces (image capture, perception output, control command issuance, actuation acknowledgment) to compute per-cycle end-to-end (E2E) latency. Technically, sub-millisecond clock synchronization across nodes is vital for isolating these one-way delay measurements from the 1--2 ms baseline network RTT. The monitoring layer also records (ii) CPU utilization on each robot node via periodic sampling at a fixed rate, and (iii) optional network statistics, including induced packet loss, injected delay, and configured loss parameters, depending on the scenario. Table~\ref{tab:testbed} summarizes the testbed hardware, software stack, and network characteristics.

%================================
\section{Execution Modes and QoS Model}\label{sec3}
%================================
\subsection{Execution Modes}
The testbed supports three execution modes that differ in the placement of the perception task T2 within the sensing--perception--planning--control pipeline. Sensing (T1) remains anchored to R1, while planning and control (T3, T4) remain anchored to R2 across all modes

\textbf{Mode 1 -- Local (LOC):}
All pipeline tasks T1--T4 execute on the robot side. T1 and T2 run on R1, while T3 and T4 run on R2; the edge node collects telemetry only. This configuration removes network delay from the control loop but results in higher CPU utilization on the robots, particularly on R1, which hosts the perception workload.

\textbf{Mode 2 -- Static Offload (SO):}
Perception (T2) is permanently offloaded to the edge node~E, while T1 remains on R1 and T3--T4 on R2. R1 streams images to~E, which returns target information to R2 for subsequent planning and control. The placement is fixed and does not react to QoS variations. This mode reduces computational load on R1 at the cost of increased network constraints and contention on~E.

\textbf{Mode 3 -- QoS-Aware Adaptive Task Placement (ATP):}
T2 can execute either locally on R1 or remotely on~E. A runtime controller observes QoS metrics and switches T2 between local and edge execution when conditions cross predefined thresholds. This mode aims to balance robot CPU utilization and control-loop latency by offloading perception to the edge when R1 is CPU-stressed and onloading it back to R1, under QoS-aware control, when network or edge conditions degrade and sufficient local headroom is available.

\subsection{QoS Metrics}
The QoS evaluation focuses on the closed-loop behavior of the visual servoing pipeline, from image capture at T1 to actuation at T4. To capture both the timing properties of the control loop and the resource usage across nodes, we define four primary metrics, summarized in Table~\ref{tab:qos-metrics}. These metrics are computed per experimental run, with latency distributions reported using high-percentile values ($L_{95}$) to highlight tail behavior.

\begin{table}[!ht]
  \centering
  \caption{QoS metrics used for closed-loop evaluation.}
  \label{tab:qos-metrics}
  \begin{tabular}{@{}llll@{}}
    \toprule
    Metric & Symbol & Unit & Role \\ \midrule
    E2E latency & $L$ & ms & Per-cycle control-loop delay \\
    Deadline & $D$ & ms & Target completion time per cycle \\
    Violation rate & $V_D$ & \% & Fraction of cycles with $L > D$ \\
    Robot CPU utilization & $U_r$ & \% & Load and headroom on R1/R2 \\
    Edge CPU utilization & $U_e$ & \% & Load and contention on node~E \\ \bottomrule
  \end{tabular}
\end{table}

\subsection{QoS-Aware Decision Rule}
To elevate the theoretical rigor of the ATP mechanism, we frame the decision engine within the agentic automation paradigm for next-gen ICPS. In this context, the controller acts as an autonomous agent exhibiting Awareness (monitoring QoS metrics), Adaptability (evaluating placement conditions), and Autonomy (executing offload/onload decisions). 

Rather than relying on arbitrary boundary checks, the static thresholds utilized in our testbed represent the pre-computed bounds for safe operational states. At runtime, the controller aggregates the QoS metrics over an observation window $k$ (comprising $W$ control cycles) and executes its autonomous logic via the following ATP decision variables:

\begin{itemize}
    \item $\hat{L}$: a recent estimate of E2E latency (per-window 95th percentile of $L$ in $W$)
    \item $L_{th}$: maximum acceptable E2E latency threshold
    \item $U_{r}^{L}$: low CPU utilization threshold, below which there is sufficient headroom to onload T2 back to R1
    \item $U_{r}^{H}$: high CPU utilization threshold, above which R1 is considered CPU-stressed and a candidate for offloading
\end{itemize}

The controller uses a single latency threshold $L_{th}$ for the control loop and two CPU thresholds for R1, a low threshold $U_{r}^{L}$ and a high threshold $U_{r}^{H}$ with $U_{r}^{L} < U_{r}^{H}$. The edge node is assumed to have sufficient compute capacity and is therefore not explicitly constrained in the decision rule. The ATP controller then decides whether T2 should remain local on R1 or be offloaded to E according to the current placement:

\begin{itemize}
    \item \textbf{Prefer edge offload (T2 on E) if} $\hat{L} \ge L_{th}$ \textbf{or} $\hat{U}_{r} \ge U_{r}^{H}$ \\
    i.e., the E2E latency exceeds the configured bound, or if the robot CPU approaches its configured threshold.
    \item \textbf{Prefer local execution (T2 on R1) if} $\hat{L} \ge L_{th}$ \textbf{or} $\hat{U}_{r} \le U_{r}^{L}$, i.e., the robot CPU has fallen below a low threshold, providing sufficient headroom to onload perception back, or as a safety fallback if the edge-offloaded latency spikes due to network degradation.
\end{itemize}

Using this decision rule implements a simple hysteresis mechanism: offloading is triggered only when R1 is clearly overloaded, or latency is over the high threshold, whereas onloading back to the robot is delayed until CPU utilization drops well below the low threshold, thereby reducing oscillations between local and edge placements. The policy is deliberately kept simple and measurement-driven so that it remains practical and reproducible on real hardware, as summarized in the following pseudocode~\ref{alg:atp}.

%=============================
\section{Experimental Setup and Evaluation}\label{sec4}
%=============================
This section describes how the ATP controller is exercised on the edge-centric MRS testbed, the scenarios used to stress the system, and how latency, deadline violations, CPU utilization, and switching behavior are evaluated across the three execution modes (LOC, SO, ATP) in this study.

%-----------------------------
\subsection{Experimental Setup}
%-----------------------------

All experiments use the hardware, network, and software configuration summarized in Table~\ref{tab:testbed} and Fig.~\ref{fig:arch}. The visual servoing pipeline implements the sensing--perception--planning--control loop on the MRS testbed. The control period $P$ is configured between $20$ and $50$~ms depending on the experiment, and the per-cycle deadline is set to $D\leq P$. The ATP controller operates with a fixed observation window of $W$ control cycles; at the end of each window, windowed estimates of the E2E latency $\hat{L}$ and robot CPU utilization $\hat{U}_r$ are computed, and the placement of T2 is updated according to the decision rule of the QoS-aware ATP controller.

To stress different aspects of the system, three classes of scenarios are considered during the experiment:

\begin{itemize}
  \item \textbf{Baseline scenario:} Default CPU load on robots with low network constraints on the WiFi LAN. This approximates a well-provisioned laboratory deployment in which both local and edge execution are in good condition.

  \item \textbf{Robot CPU stress scenario:} Synthetic CPU load is injected on R1 using workload generators to emulate competing processes or additional sensing pipelines. Network conditions remain close to baseline. This scenario highlights how LOC, SO, and ATP behave when robot-side compute becomes the primary bottleneck.

  \item \textbf{Network impairment scenario:} Linux \texttt{tc netem} on the edge interface emulates industrial wireless degradation while robot CPU loads remain at baseline. To model factory-floor interference, we inject a Gaussian one-way delay ($\mu = 25$ ms, $\sigma = 5$ ms) and a uniform random packet loss model with $p_{loss} = 2\%$. This rigorously tests the ATP controller's safety fallback mechanism.
\end{itemize}

Each scenario is executed under all three execution modes (LOC, SO, ATP). For every mode--scenario combination, the system is run for a fixed number of control cycles (on the order of tens of thousands) to obtain steady-state statistics. Raw logs record timestamps at the T1 and T4 interfaces, CPU samples on R1 and R2, the active execution mode or T2 placement per window, and the network emulation parameters. Where applicable, the same random seeds and workload scripts are reused across runs to improve comparability between modes.

%-----------------------------------
\subsection{Latency and Deadline Violations}
%-----------------------------------

Latency and deadline violations are evaluated using the QoS metrics in Table~\ref{tab:qos-metrics}. For each control cycle, the E2E latency $L$ is computed as the time difference between image capture at T1 and actuation command issuance at T4. Over a given run, the empirical distribution of $L$ is summarized by:
\begin{itemize}
    \item the mean ($\mu_L$) and standard deviation ($\sigma_L$) of L,
    \item the $L_{95}$ to characterize tail behavior in $W$, and
    \item the cumulative distribution function (CDF) of $L$ for visual comparison between modes of this empirical study.
\end{itemize}

The violation rate $V_D$ denotes the fraction of cycles in which $L>D$. We report $V_D$ and $L_{95}$ across modes to evaluate the trade-off between deadline robustness and tail latency. For ATP, time-series analysis correlates $L$ and $V_D$ with placement decisions to verify that the controller correctly offloads during local slowdowns and onloads when CPU headroom and latency recover.

\begin{algorithm}[t]
\caption{Adaptive Task Placement (ATP) Policy}
\label{alg:atp}
\begin{algorithmic}[1]
\State \textbf{Inputs:} window length $W$, latency threshold $L_{\text{th}}$, robot CPU thresholds $U_r^{H/L}$, minimum dwell windows $N_{\min}$
\State \textbf{State:} current placement $m \in \{\textsc{LOC}, \textsc{SO}\}$, dwell counter $n_{\text{dwell}} \gets 0$
\For{each window $k = 1,2,\dots$}
  \State Run loop for $W$ cycles under placement $m$
  \State Compute windowed estimates $\hat{L}_k$ and $\hat{U}_{r,k}$
  \If{$n_{\text{dwell}} < N_{\min}$}
    \State $n_{\text{dwell}} \gets n_{\text{dwell}} + 1$
    \State \textbf{continue} \Comment{enforce minimum dwell time}
  \EndIf
  \If{$m = \textsc{LOC}$}
    \If{$\hat{L}_k \geq L_{\text{th}}$ \textbf{or} $\hat{U}_{r,k} \geq U_r^{H}$}
      \State $m \gets \textsc{SO}$; $n_{\text{dwell}} \gets 0$ \Comment{offload T2 to E}
    \EndIf
  \ElsIf{$m = \textsc{SO}$}
    \If{$\hat{L}_k \geq L_{\text{th}}$ \textbf{or} $\hat{U}_{r,k} \leq U_r^{L}$}
      \State $m \gets \textsc{LOC}$; $n_{\text{dwell}} \gets 0$ \Comment{onload T2 back / safety fallback}
    \EndIf
  \EndIf
\EndFor
\end{algorithmic}
\end{algorithm}

%---------------------------------------------------
\subsection{CPU Utilization and Switching Behavior}
%---------------------------------------------------

\begin{figure*}[t]
  \centering

  % ---------- Top row ----------
  \begin{subfigure}[t]{0.48\textwidth}
    \centering
    \includegraphics[width=0.7\linewidth]{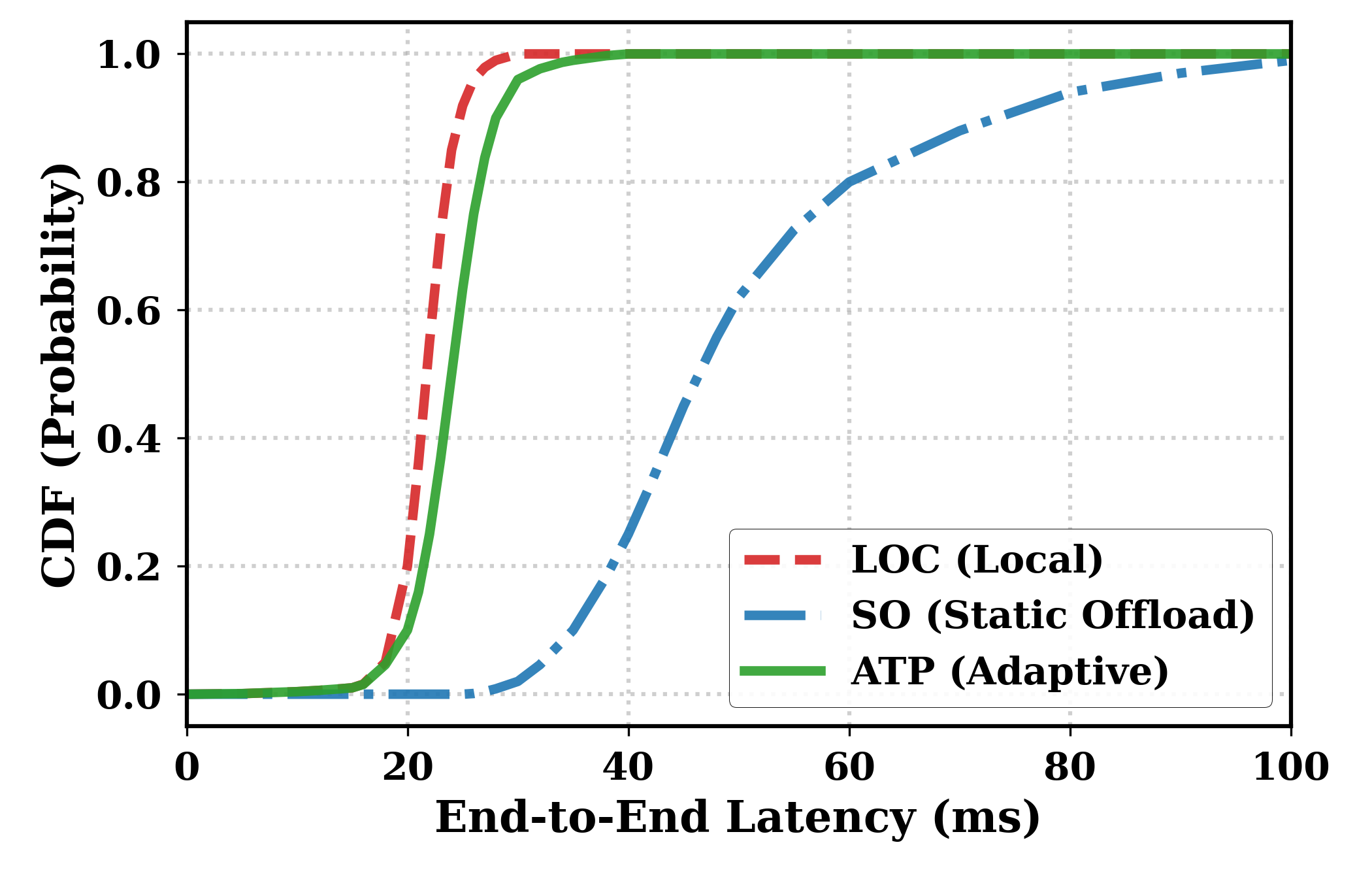}%
    \subcaption{End-to-end latency distribution for LOC, SO, and ATP.}
    \label{fig:2a}
  \end{subfigure}
  \hfill
  \begin{subfigure}[t]{0.48\textwidth}
    \centering
    \includegraphics[width=0.7\linewidth]{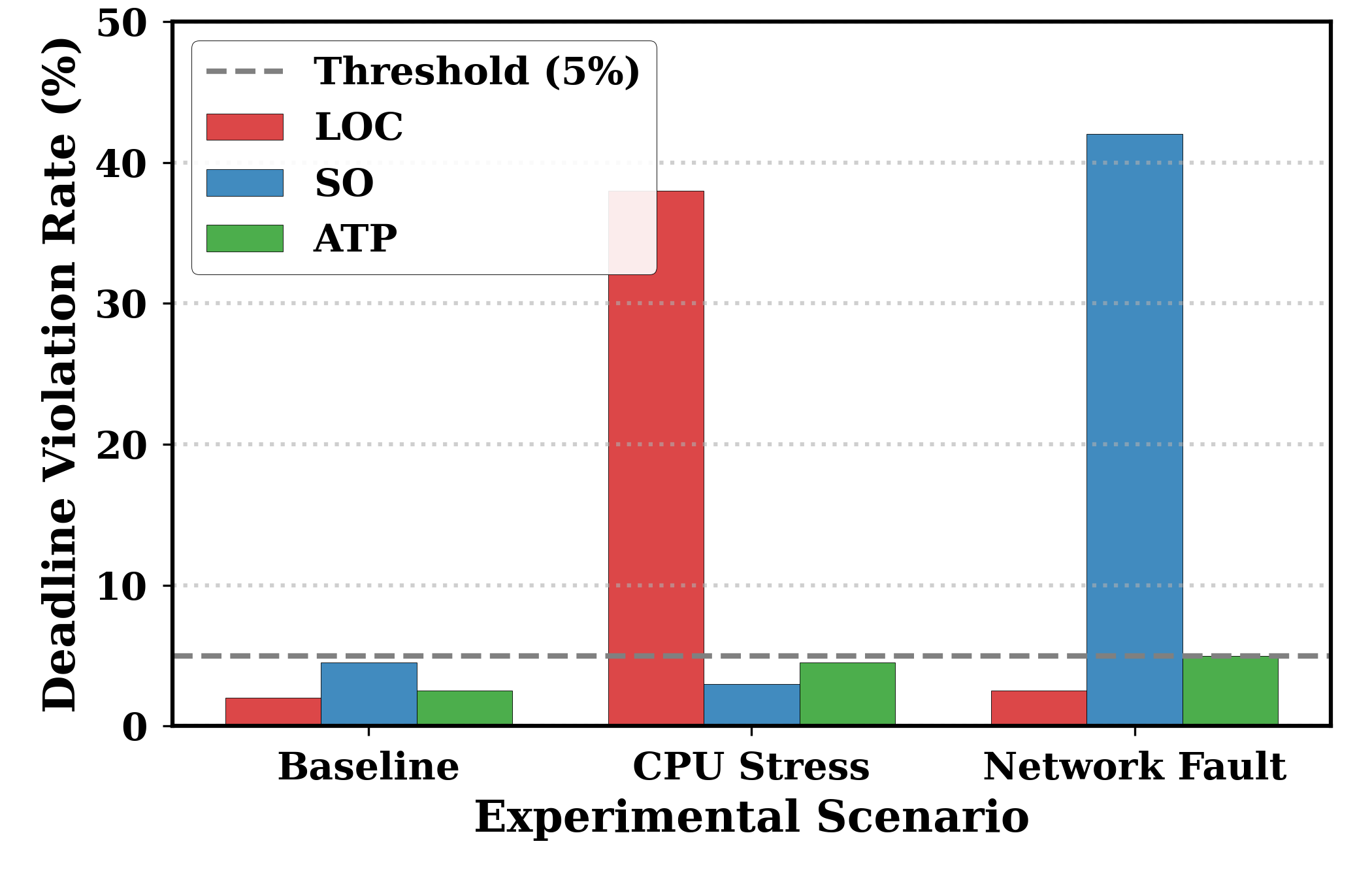}%
    \subcaption{Deadline violation rates ($V_D$) across experimental scenarios.}
    \label{fig:2b}
  \end{subfigure}

  \vspace{0.3cm}

  % ---------- Bottom row ----------
  \begin{subfigure}[t]{0.48\textwidth}
    \centering
    \includegraphics[width=0.6\linewidth]{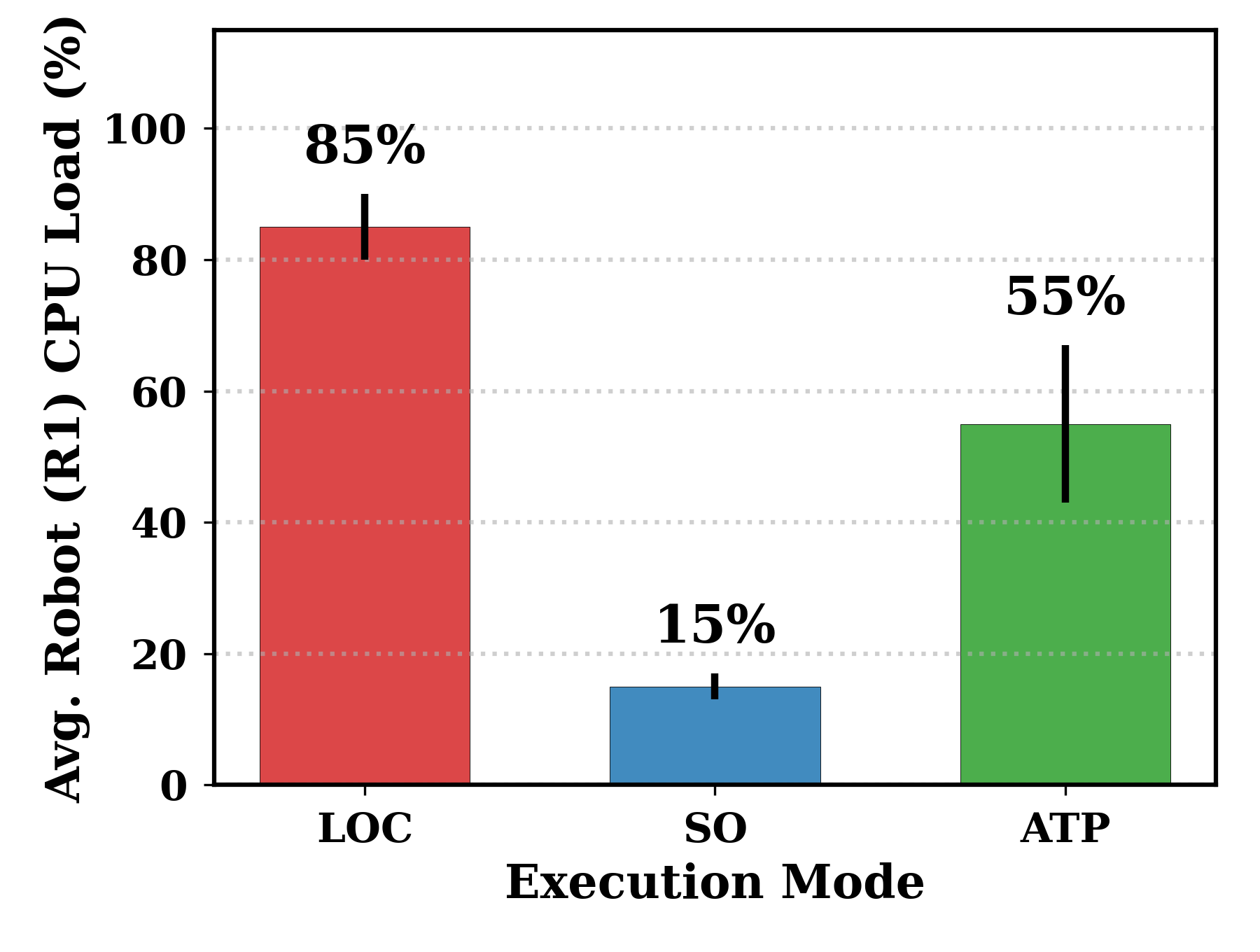}%
    \subcaption{Average CPU utilization ($U_r$) on R1 for each execution mode.}
    \label{fig:2c}
  \end{subfigure}
  \hfill
  \begin{subfigure}[t]{0.48\textwidth}
    \centering
    \includegraphics[width=\linewidth]{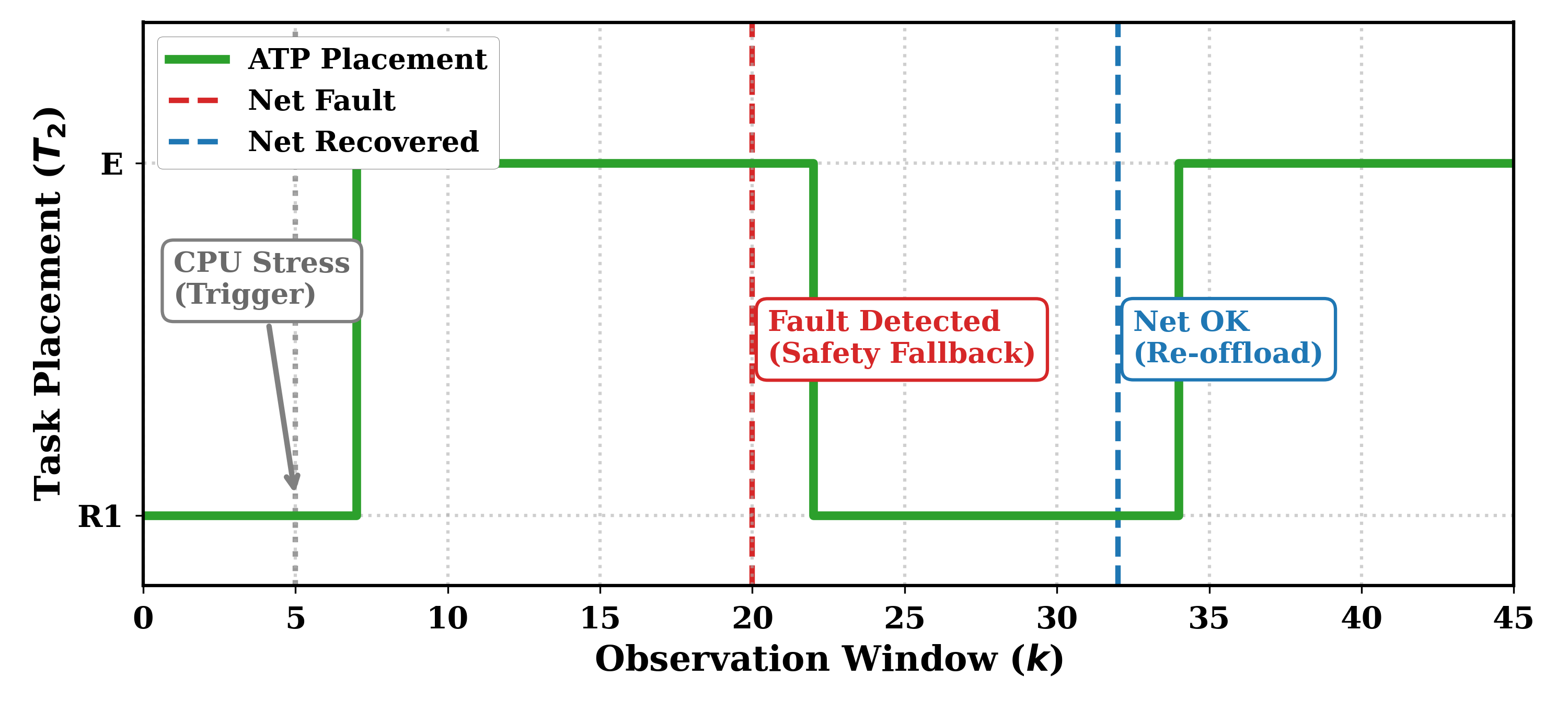}%
    \subcaption{Multi-stage ATP switching T2 placement over decision windows.}
    \label{fig:2d}
  \end{subfigure}

  \caption{QoS metrics and controller behavior for the three execution modes (LOC, SO, ATP), including latency distributions, deadline violations, R1 CPU resource utilization, and ATP placement dynamics across scenarios.}
  \label{fig:2}
\end{figure*}

For every control cycle, the instantaneous CPU utilization of R1 and R2 is sampled using standard system tools and recorded in the raw logs. After $W$ cycles of the experiment, the controller computes windowed estimates $U_{r}$, where applicable, and $U_{e}$ as the average CPU utilization on the robot and edge nodes, respectively. These estimates are used both as inputs to the ATP decision rule and as metrics for post-hoc evaluation.

For each scenario and mode, CPU behavior is summarized using window-level statistics:
\begin{itemize}
    \item the $\mu_{U_{r}}$ and $\sigma_{U_{r}}$ characterizing the average compute load and its variability over the experiment, and
    \item the empirical distribution of ${U}_{r}$ across windows, highlighting how often the robot CPU operates near saturation under LOC, SO, and ATP.
\end{itemize}

Switching behavior is examined at the granularity of the ATP observation window. For each run, the active execution mode per window is extracted from the logs, and the following indicators are computed:
\begin{itemize}
    \item the fraction of windows spent in each placement (LOC-like vs.\ SO-like) under ATP,
    \item the number of mode switches per $N$ windows as a measure of controller reactivity, and
    \item the distribution of dwell times, i.e., consecutive windows spent in the same mode, as an indicator of stability.
\end{itemize}

These CPU and switching metrics are analyzed jointly with the latency and deadline statistics. In particular, we inspect how increases in $\hat{U}_{r}$ correlate with the ATP controller migrating T2 away from R1, and whether reductions in R1 load lead to sustained periods of local execution without excessive switching. This allows us to assess whether pilot ATP achieves a desirable trade-off between responsiveness to CPU stress and stable placement decisions across the three scenario classes.

\section{Results and Discussion}\label{sec5}
%========================
This section interprets the evaluation results and findings of this study, focusing on how LOC, SO, and ATP differ in latency, deadline robustness, CPU utilization, and switching behavior across the three scenarios. The aim is to highlight key trends that emerge from the empirical study and to assess when the ATP controller provides measurable QoS benefits over static execution. Unless otherwise stated, results are averaged over multiple runs per mode--scenario pair. Fig.~\ref{fig:2} summarizes the key QoS outcomes in terms of latency distributions, deadline violations, robot CPU utilization, and ATP switching dynamics. Overall, the findings offer initial insight into the conditions under which adaptive task placement is most effective and motivate directions for future work.

%----------------------------
\subsection{Latency and CPU Utilization}
%----------------------------

The fundamental trade-off in edge robotics, balancing execution speed with local resource consumption, is illustrated in Fig.~\ref {fig:2}. Fig.~\ref{fig:2a} presents the CDF of end-to-end latency. The LOC mode (red dashed line) demonstrates the steepest rise, indicating low and consistent latency as it avoids network transmission overhead. Crucially, the proposed ATP mode (green solid line) closely mirrors the performance of LOC, achieving a near-identical latency profile. In contrast, the SO mode (blue dash-dot line) exhibits a significant ``long tail'' and higher overall latency, attributable to network round-trip times (RTT) and serialization overheads. This confirms that ATP successfully identifies when network conditions are suboptimal or when local execution is preferable, preventing the latency penalty observed in static offloading.

Fig.~\ref{fig:2c} details the average CPU Load on R1:
\begin{itemize}
    \item \textbf{LOC:} As expected, purely local execution incurs the highest computational cost, saturating the robot's CPU at approximately $85\%$. This leaves little headroom for other critical background processes onboard.
    \item \textbf{SO:} Static offloading minimizes local strain, reducing utilization to $15\%$, but at the cost of the latency penalties observed in Fig.~\ref{fig:2a} due to network communication.
    \item \textbf{ATP:} The adaptive approach strikes an optimal balance. By dynamically offloading only when beneficial, ATP maintains an average CPU load of $55\%$. This represents a $30\%$ reduction in local resource usage compared to LOC, ensuring the robot remains critically responsive without over-relying on the edge nodes, even the cloud server.
\end{itemize}

\subsection{System Robustness and Deadline Compliance}
The contribution is the resilience to environmental variances. Fig.~\ref{fig:2b} compares the Deadline Violation Rate ($V_D$) across scenarios: \lq Baseline\rq~(nominal), \lq CPU Stress\rq~(local resource contention), and \lq Network Fault\rq~(packet loss/bandwidth reduction). A dashed line at $5\%$ indicates the maximum acceptable violation rate for the control loop.

\begin{enumerate}
    \item \textbf{Baseline:} All three modes perform well, staying below the $5\%$ threshold (only focus on R1).
    \item \textbf{CPU Stress:} When R1 is subjected to heavy computational load, the LOC mode fails catastrophically, with violation rates spiking to nearly $40\%$. Because LOC cannot offload, the tasks queue up and miss deadlines. SO and ATP remain stable as they leverage the edge.
    \item \textbf{Network Fault:} Conversely, when the network quality degrades, the SO mode suffers, exceeding a $40\%$ violation rate due to transmission delays. LOC remains stable as it does not rely on network communication.
\end{enumerate}

The ATP mechanism is the only mode that remains robust across all scenarios. In the \lq CPU Stress\rq~condition, it behaves like SO; in the \lq Network Fault\rq~condition, it behaves like LOC. Consequently, ATP consistently maintains the deadline violation rate below the critical $5\%$ threshold, validating its ability and resilience to adapt to complex dynamic constraints. Unlike hard real-time motion control, this soft real-time visual servoing task tolerates minor jitter; thus, we establish a 5\% deadline violation threshold ($V_D \le 5\%$), aligning with standard Service Level Agreements (SLAs) for edge-assisted machine vision \cite{julia_tsn_2025} to prevent manipulator destabilization.

\subsection{Dynamic Switching Behavior}
To understand the decision-making process of the controller, Fig.~\ref{fig:2d} visualizes the task -- T2 over a time-series observation window ($k$). The trace begins with the system in a low-load state. At $k=5$ (CPU Stress Trigger), the local CPU load increases. The ATP controller detects this contention and switches placement from R1 to E to alleviate pressure on R1.

Later, at $k=20$ (Network Fault), a fault is injected into the communication channel. The controller detects the degradation in offloading performance and executes a \lq Safety Fallback\rq, reverting execution to R1. This prevents the high deadline violations seen in the SO mode. Finally, at $k=32$ (Network Recovered), once the network fault clears and stability is confirmed, the system executes a \lq re-offload\rq~back to E to reduce local stress (during triggered \lq CPU Stress\rq~state). This multi-stage switching demonstrates the stability and responsiveness of the ATP controller, confirming it acts as a fail-safe mechanism for safety-critical robotic workloads.

%================
\section{Conclusions and Future Work}\label{sec6}
%================
This paper presented an edge-centric MRS testbed that couples two heterogeneous Raspberry Pi-based robots with an edge node executing a camera-to-manipulator visual servoing pipeline. Three execution modes were implemented and evaluated: fully local on-robot execution (LOC), static edge offloading (SO), and a QoS-aware ATP controller that relocates the task (T2) between R1 and E. The study defined a simple QoS model centered on E2E latency, deadline violations, and robot CPU utilization. It instantiated ATP as a feedback-driven controller with a single latency threshold and CPU hysteresis.

Experimental results show that static edge offloading can effectively reduce robot CPU load but is fragile under variable network conditions, leading to amplified tail latency and increased deadline violations. 
LOC remains robust to network impairments but offers limited CPU headroom on resource-constrained robots. 
In contrast, ATP improved deadline robustness while preserving explicit CPU headroom on R1 under various network conditions. 
These findings position ATP as a practical control primitive for future agentic automation in MRS deployments, but limited to complex network conditions.
Overall, this yields concrete design guidelines for next-generation task placement mechanisms, which solidify CER as the architectural reference for networked robotics.

Future work will extend the testbed in several directions. First, the QoS-aware ATP logic will be generalized to richer task graphs in multi-robot fleets, enabling dynamic placement over full perception-planning DAGs. Second, more expressive optimization frameworks, such as multi-objective formulations and learning-based policies, will be explored on top of the same QoS metrics as a co-design mechanism. Finally, this setup will be integrated into a broader cloud-fog stack (wide-area links and cloud services) to study hierarchical orchestration, alongside the investigation of additional QoS dimensions (e.g., Age-of-Information, safety margins) and alternative network substrates like AI-native TSN~or~5G--URLLC.

\bibliographystyle{IEEEtran}
\bibliography{refs}

\end{document}